\documentstyle[prb,twocolumn,aps,epsf]{revtex}
\newcommand{\bm}[1]{\mbox{\boldmath$#1$}}
\begin{document} \draft 
\twocolumn[
\title{Spin-Gap Phases in Tomonaga-Luttinger Liquids}
\author{Masaaki Nakamura\cite{email1}}
\address{
Institute for Solid State Physics, University of Tokyo,
Roppongi, Tokyo 106-8666, Japan}
\author{Atsuhiro Kitazawa\cite{email2} and Kiyohide Nomura\cite{email3}}
\address{
Department of Physics, Kyushu University, Fukuoka 812-8581, Japan}
\date{December 14, 1998}
\maketitle 
\begin{abstract}
\widetext\leftskip=0.10753\textwidth \rightskip\leftskip

 We give the details of the analysis for critical properties of spin-gap
 phases in one-dimensional lattice electron models.  In the
 Tomonaga-Luttinger (TL) liquid theory, the spin-gap instability occurs
 when the backward scattering changes from repulsive to attractive.
 This transition point is shown to be equivalent to that of the
 level-crossing of the singlet and the triplet excitation spectra, using
 the $c=1$ conformal field theory and the renormalization group. Based
 on this notion, the transition point between the TL liquid and the
 spin-gap phases can be determined with high-accuracy from the numerical
 data of finite-size clusters.  We also discuss the boundary conditions
 and discrete symmetries to extract these excitation spectra.  This
 technique is applied to the extended Hubbard model, the $t$-$J$ model,
 and the $t$-$J$-$J'$ model, and their phase diagrams are obtained. We
 also discuss the relation between our results and analytical solutions
 in weak-coupling and low-density limits.

\end{abstract}

] \narrowtext

\section{Introduction}\label{sec:intro}

Spin-gap phases of one-dimensional (1D) electron systems have been
studied for long times. This research has been motivated by the phase
transitions in 1D organic conductors. In the past decade, the discovery
of high-$T_{\rm c}$ superconductivity strongly stimulated this study.

The spin-gap transition in 1D lattice models had been mainly analyzed by
two approaches: The one is the weak coupling theory based on the
bosonization theory and renormalization group. The other is numerical
calculation in finite-size systems which is free from approximation. In
the former scheme, the existence of the gap is argued by investigating
the backward scattering effect on the fixed point, but the validity of
the result is ensured only in the weak coupling limit. On the other
hand, in numerical calculation, the analysis is done by a direct
evaluation of the gap and the finite-size scaling method. In this
approach, a singular behavior of the gap near the critical point makes
it difficult to make out the instability.

In order to illustrate the difficulty in the determination of the phase
boundary, let us consider the Hubbard model,
\begin{equation}
 {\cal H}_{\rm HM}=
 -t\sum_{is}(c^{\dag}_{is} c_{i+1,s}+\mbox{H.c})
 +U\sum_i n_{i\uparrow}n_{i\downarrow}.
\end{equation}
This model has a spin gap for $U<0$.  According to the Bethe-ansatz
result for the charge gap at half-filling\cite{Ovchinnikov} combining a
canonical transformation \cite{Shiba}, we can obtain the asymptotic
behavior of the spin gap near the critical point $U=0$ as
\begin{equation}
 \Delta E\sim \sqrt{2t|U|}{\rm e}^{-\pi t/|U|}.
\label{eqn:gap_Hubbard}
\end{equation}
Since the gap opens slowly near the critical point, it is very difficult
to find the critical point using conventional finite-size scaling
method.

In this paper, we give a remedy for this
problem\cite{Nakamura-N-K,Nakamura}.  The many-body problem is often
simplified by using the notion of universality. Generally, 1D electron
systems belong to the universality class of Tomonaga-Luttinger (TL)
liquids\cite{Tomonaga-L,Haldane,Emery,Solyom,Voit} which are
characterized by gapless charge and spin excitations and power-law decay
of correlation functions. This behavior can be described by the
bosonization theory or the $c=1$ conformal field theory (CFT). In this
scheme, the phase transition to the spin-gap phase is understood as an
instability caused by the backward scattering process using the
renormalization group technique\cite{Manyhard-S}, and a spin gap opens
when the backward scattering turns from repulsive to attractive.  This
transition point is equivalent to the level-crossing of the
singlet-triplet excitation
spectra\cite{Affleck-G-S-Z,Ziman-S,Okamoto-N}, by taking account of the
logarithmic corrections originated from the backward scattering.

In this paper, we will analyze the following models based on this
notion. The first example is the extended Hubbard model which is
given by
\begin{equation}
 {\cal H}_{\rm EHM}={\cal H}_{\rm HM}+V\sum_i n_{i}n_{i+1}. 
\label{eqn:EHM}
\end{equation}
For the study of spin-gap transitions, this model has been analyzed by
the {\it g-ology} for weak coupling region\cite{Emery,Solyom}. The
numerical calculation was performed by the exact diagonalization with
finite-size scaling method\cite{Penc-M,Sano-O,Lin-G-C-F-G}.  However,
the spin-gap phase boundary has not been clarified.

The next example is the $t$-$J$ model described by
\begin{eqnarray}
{\cal H}_{t\mbox{-}J}&=&
 -t\sum_{is}(\tilde{c}^{\dag}_{is} \tilde{c}_{i+1,s}+\mbox{H.c.})\nonumber\\
 &&+J\sum_{i}(\bm{S}_i\cdot\bm{S}_{i+1}-n_i n_{i+1}/4),
 \label{eqn:t-J}
\end{eqnarray} 
where $\tilde{c}_{is}=c_{is}(1-n_{i,-s})$. This model is obtained by
doping holes in the Heisenberg spin chain. For this model, the weak
coupling treatment is difficult due to this strong coupling constraint,
however, the universality class of this model is known as TL liquids,
from the analysis for the exactly solvable cases at $J/t=0$ (spinless
fermion) and $J/t=2$ (super-symmetric
point)\cite{Bares-B-O,Kawakami-Y90b}. The remaining region was analyzed
using the exact diagonalization by Ogata {\it et al.}\cite{Ogata-L-S-A}. 
Their phase diagram shows the enhancement of the superconducting
correlation ($K_{\rho} > 1$) and the phase separation
($K_{\rho}\rightarrow\infty$) for the large $J/t$ region. According to
their result, the spin-gap phase does not exist except for the low
density region. Variational approaches also played roles
in the analysis for this model\cite{Hellberg-M,Chen-L,Yokoyama-O}, but
could not establish clear solution for the spin-gap phase.

Extensions of the $t$-$J$ model are also considered by many researchers
\cite{Sano-T,Ogata-L-R,Imada,Dagotto-R,Troyer-T-R-R-D,Ammon-T-T}.
In spin systems, a spin gap opens by the effect of frustration or
dimerization. Metallic spin-gap phases can be generated by doping holes
in these spin systems. In this paper, we concentrate our attention on
the $t$-$J$-$J'$ model which includes the effect of
frustration\cite{Sano-T,Ogata-L-R}:
\begin{eqnarray}
{\cal H}_{t\mbox{-}J\mbox{-}J'}&=&
{\cal H}_{t\mbox{-}J}+J'\sum_{i}(\bm{S}_i\cdot\bm{S}_{i+2}-n_i n_{i+2}/4).
\label{eqn:t-J-J'}
\end{eqnarray}
We introduce a parameter $\alpha$ for the strength of the frustration
given by $\alpha\equiv J'/J$. At half-filling ($n=1$), this model
becomes an $S=1/2$ frustrated spin chain. In this case, the ground state
at $\alpha=1/2$ is the two-fold degenerate dimer state with a spin gap,
and the ground state energy density is
$-3/4J$\cite{Majumder-G,Shastry-S,Broek}. The fluid-dimer transition
occurs at $\alpha_{\rm c}=0.2411$\cite{Okamoto-N}. Upon doping of holes,
the system may become metallic, and the spin gap is reduced\cite{Sano-T}
but persists for the finite doping. The phase diagram of this model for
$n\neq 1$ at $\alpha=1/2$, using the exact diagonalization, was obtained
by Ogata, Luchini, and Rice\cite{Ogata-L-R}, but the phase boundary of
the spin-gap phase was also remained to be ambiguous.

This paper is organized as follows. In Sec.\ref{sec:formalism}, we
discuss, based on the continuum field theory, that the level crossing of
singlet and triplet excitation spectra gives the critical point of the
spin-gap transition. In Sec.\ref{sec:symmetry}, we consider boundary
conditions for the unique ground state, and discrete symmetries of wave
functions to identify the energy spectra observed in our analysis. In
Sec.\ref{sec:diagrams}, we analyze representative models introduced
above, and clarify the spin-gap region in the phase diagrams, and check
the consistency of our argument.  Finally, in Sec.\ref{sec:summary}, we
present our conclusions.

The paper also contains three Appendices.  The first shows the relation
among the different notations for the quantum numbers. The second is
derivation for the logarithmic corrections.  The third explains the
calculation in two-electron systems.

\section{Continuum Field Theory}
\label{sec:formalism}
\subsection{Effective Hamiltonian}
Let us start our argument from the Abelian bosonization theory of
electrons\cite{Haldane,Emery,Solyom,Voit,Schulz}. The low-energy
excitations are described by continuous fermion fields which are defined
by
\begin{equation}
 c_{j,s}\rightarrow \psi_{{\rm L},s}(x)+\psi_{{\rm R},s}(x)
\end{equation}
The boson representation of the fermion operator is
\begin{equation}
 \psi_{r,s}(x)=\frac{1}{\sqrt{2\pi\alpha}}
  {\rm e}^{{\rm i} r k_{\rm F} x}{\rm e}^{{\rm i}/\sqrt{2}\cdot
 [r(\phi_{\rho}+s\phi_{\sigma})-\theta_{\rho}-s\theta_{\sigma}]},
  \label{eqn:fermion_op}
\end{equation}
where $\alpha$ is a short-distance cutoff. $r={\rm R},{\rm L}$ and
$s=\uparrow, \downarrow$ refer to $+,-$ in that order. The phase fields
are defined as
\begin{mathletters}
\label{eqn:phase_fields}
\begin{eqnarray}
\phi_{\nu}(x)
  &=&-\frac{{\rm i}\pi}{L}\sum_{p\neq 0}A_p(x)
  \left[\nu_{\rm R}(p) + \nu_{\rm L}(p)\right]
  -\frac{\sqrt{2}\pi x}{L}\hat{n}_{\nu},\\
\theta_{\nu}(x)
  &=&+\frac{{\rm i}\pi}{L}\sum_{p\neq 0}A_p(x)
  \left[\nu_{\rm R}(p) - \nu_{\rm L}(p)\right]
  +\frac{\sqrt{2}\pi x}{L}\hat{m}_{\nu},
\end{eqnarray}
\end{mathletters}
where $A_p(x)\equiv\frac{1}{p}{\rm e}^{-{\rm i}\alpha|p|/2-{\rm i}px}$,
and $\nu_{r}$ is the charge ($\nu=\rho$) or the spin ($\nu=\sigma$)
density operator. These phase fields satisfy the relation
$\left[\phi_{\nu}(x),\theta_{\nu}(x')\right]=-{\rm i}\pi
{\rm\,sign}(x-x')/2$. 

Using above relations, effective Hamiltonian of a 1D electron
system is described by the U(1) Gaussian model (charge part) and the
SU(2) sine-Gordon model (spin part),
\begin{equation}
 {\cal H}={\cal H}_{\rho} + {\cal H}_{\sigma}
  +\frac{2 g_{1\perp}}{(2\pi\alpha)^2}
 \int_0^L {\rm d}x \cos(\sqrt{8}\phi_{\sigma})
 \label{eqn:effHam}.
\end{equation}
Here $g_{1\perp}$ is the backward
scattering amplitude and for $\nu=\rho,\sigma$
\begin{equation}
  {\cal H}_{\nu}=\frac{v_{\nu}}{2\pi}\int_0^L {\rm d}x
  \left[K_{\nu}(\partial_x \theta_{\nu})^2
  +K_{\nu}^{-1}(\partial_x \phi_{\nu})^2\right]
\label{eqn:Gaussian},
\end{equation}
where $v_{\nu}$ and $K_{\nu}$ are the velocity and the Gaussian
coupling, respectively, for the charge ($\nu=\rho$) and the spin
($\nu=\sigma$) sectors. In the TL phase ($g_{1\perp}>0$), the parameters
$K_{\sigma}$ and $g_{1\perp}$ are renormalized as $K^{*}_{\sigma}=1$ and
$g_{1\perp}^{*}=0$, reflecting the SU(2) symmetry.

The phase fields defined in eqs.(\ref{eqn:phase_fields}) satisfy the
following boundary conditions,
\begin{mathletters}
\begin{eqnarray}
 \phi_{\nu}(x+L)&=&\phi_{\nu}(x)-\sqrt{2}\pi n_{\nu},\\
 \theta_{\nu}(x+L)&=&\theta_{\nu}(x)+\sqrt{2}\pi m_{\nu}.
\end{eqnarray}\label{eqn:BC_phase}
\end{mathletters}
The quantum numbers $m_{\nu}$ and $n_{\nu}$ are defined by the eigen
values of the total number operators $\hat{N}_{r,s}$ (measured with
respect to the ground state) for right and left going particles ($r={\rm
R},{\rm L}$) of spin $s$
\begin{mathletters}
\begin{eqnarray}
 n_{\nu}&=&
  [(N_{{\rm R}\uparrow}+N_{{\rm L}\uparrow})
\pm (N_{{\rm R}\downarrow}+N_{{\rm L}\downarrow})]/2,\\
 m_{\nu}&=&
  [(N_{{\rm R}\uparrow}-N_{{\rm L}\uparrow})
\pm (N_{{\rm R}\downarrow}-N_{{\rm L}\downarrow})]/2.
\end{eqnarray}\label{eqn:q_num}
\end{mathletters}
Here the upper and lower sign refer to charge ($\nu=\rho$) and spin
($\nu=\sigma$) degrees of freedoms, respectively. Thus $n_{\nu}$ denotes
excitations involving the variation of particles numbers and $m_{\nu}$
indicate current excitations. If we require $N_{r,s}$ to be an integer,
the possible value of the quantum numbers are restricted as
\begin{equation}
 (-1)^{m_{\rho}\pm m_{\sigma}}=(-1)^{n_{\rho}\pm n_{\sigma}}.
  \label{eqn:sel_rule}
\end{equation}
This is the selection rule for the quantum
numbers\cite{Haldane,Woynarovich}.

\subsection{Excitation Spectra and Boundary Conditions}
First, we consider the excitation spectra for $g_{1\perp}=0$ case. If
the system is periodic, and has unique ground state, the ground state
energy of the system with length $L$ is given
by\cite{Blote-C-N}
\begin{equation}
 E_0(L)=L\epsilon_0-\frac{\pi(v_{\rho}+v_{\sigma})}{6L}c,
\end{equation}
where the central charge $c$ characterizes the universality class of the
model. The finite-size corrections for the excitation energy and
momentum of the system are described by\cite{Cardy84,Haldane}
\begin{eqnarray}
 E-E_0&=&
  \frac{2\pi v_{\rho}}{L}x_{\rho}+\frac{2\pi v_{\sigma}}{L}x_{\sigma},
  \label{eqn:energy}\\
 P-P_0&=&\frac{2\pi}{L}(s_{\rho}+s_{\sigma})+2m_{\rho}k_{\rm F},
  \label{eqn:momentum}
\end{eqnarray}
where $k_{\rm F}=\pi N/2L$ is the Fermi wave number. $x_\nu=\Delta_\nu^+
+ \Delta_\nu^-, s_\nu=\Delta_\nu^+ - \Delta_\nu^-$ are the scaling
dimension and the conformal spin, respectively, where the conformal
weights for each sector are given by
\begin{equation}
\Delta_{\nu}^{\pm}=\frac{1}{2}
    \left(\sqrt{\frac{K_{\nu}}{2}}m_{\nu}
     \pm\frac{n_{\nu}}{\sqrt{2K_{\nu}}}\right)^2+n_{\nu}^{\pm}.
\label{eqn:weight}
\end{equation}
Here the integer $n_{\nu}^{\pm}$ denote descendant fields which describe
particle-hole excitations near the Fermi points.  The scaling dimensions
are related to the critical exponents for the correlation functions as
\begin{equation}
 \langle{\cal O}_i(r){\cal O}_i(r')\rangle
 \sim |r-r'|^{-2(x_{\rho i}+x_{\sigma i})}.
\end{equation}
Therefore, there is one to one correspondence between the excitation
spectra and the operators. The operators correspond to the excited
states are given by
\begin{equation}
{\cal O}_{m_{\rho},m_{\sigma},n_{\rho},n_{\sigma}}
\propto{\rm e}^{{\rm i}\sqrt{2}(m_{\rho}\phi_{\rho}+m_{\sigma}\phi_{\sigma}
   +n_{\rho}\theta_{\rho}+n_{\sigma}\theta_{\sigma})},
\end{equation}
or their linear combinations.  From eqs.(\ref{eqn:fermion_op}) and
(\ref{eqn:BC_phase}), the Fermi operator takes the following boundary
conditions depending on the excited states:
\begin{equation}
 \psi_{r,s}(x+L)=\psi_{r,s}(x)
{\rm e}^{{\rm i}\pi(m_{\rho}+m_{\sigma}+n_{\rho}+n_{\sigma})}.
\label{eqn:BC_fermion}
\end{equation}
This means that the excited states given by arbitrary combination of
quantum numbers are realized by changing the boundary conditions, while,
for fixed boundary conditions, the possible excited states are
restricted by the selection rule (\ref{eqn:sel_rule}).


The excitation spectra on which we will turn our attention can be
obtained based on the operators for the charge-density-wave (CDW) and
the spin-density-wave (SDW):
\begin{mathletters}
\begin{eqnarray}
{\cal O}_{\rm CDW}
&=&\psi^{\dag}_{{\rm L}\uparrow}\psi_{{\rm R}\uparrow}
  +\psi^{\dag}_{{\rm L}\downarrow}\psi_{{\rm R}\downarrow}\nonumber\\
&=&\frac{1}{\pi\alpha}
 \exp({\rm i}2k_{\rm F}x+{\rm i}\sqrt{2}\phi_{\rho})
 \cos(\sqrt{2}\phi_{\sigma}),\\
{\cal O}_{{\rm SDW},z}
&=&\psi^{\dag}_{{\rm L}\uparrow}\psi_{{\rm R}\uparrow}
  -\psi^{\dag}_{{\rm L}\downarrow}\psi_{{\rm R}\downarrow}\nonumber\\
&=&\frac{{\rm i}}{\pi\alpha}
 \exp({\rm i}2k_{\rm F}x+{\rm i}\sqrt{2}\phi_{\rho})
 \sin(\sqrt{2}\phi_{\sigma}),\\
{\cal O}_{{\rm SDW},+}
&=&\psi^{\dag}_{{\rm L}\uparrow}\psi_{{\rm R}\downarrow}\nonumber\\
&=&\frac{1}{2\pi\alpha}
 \exp({\rm i}2k_{\rm F}x+{\rm i}\sqrt{2}\phi_{\rho})
 \exp(+{\rm i}\sqrt{2}\theta_{\sigma}).
\end{eqnarray}
\label{eqn:operators}
\end{mathletters}
These excitation spectra consist of the charge part which carries the
momentum $2k_{\rm F}$, and the spin part which forms singlet
($\sqrt{2}\cos\sqrt{2}\phi_{\sigma}$) and triplet
($\sqrt{2}\sin\sqrt{2}\phi_{\sigma}, \exp(\pm{\rm
i}\sqrt{2}\theta_{\sigma})$) states. Note that the spin part of the
singlet and the triplet superconducting operators (SS, TS) are obtained
with $k_{\rm F}=0$ and replacing $\phi_\rho\rightarrow\theta_{\rho}$.

If charge-spin separation occurs, the spin excitations in
eqs.(\ref{eqn:operators}) ($m_{\sigma}=1$ or $n_{\sigma}=1$, otherwise
$=0$) can be extracted by using anti-periodic boundary conditions
following eq.(\ref{eqn:BC_fermion}). In the continuum field theory based
on the TL model, the dispersion relation is approximated by linearized
one, so that the deviation from the approximated dispersion become
smaller if the excitation energies become lower by eliminating the
charge excitations. Therefore, the precision of the analysis is enhanced
by twisting the boundary conditions.

The twisted boundary conditions are also important in identification of
excitation spectra.  Under anti-periodic boundary conditions, the
momenta of these states are reduced to zero.  Then we can define the
parity transformation to classify these spectra. Although the
space-inversion operator and translation operator do not commute, we can 
classify these spectra simultaneously by wave numbers and parities, if
the wave number $k$ takes $0$ or $\pi$.  From eq.(\ref{eqn:fermion_op}),
the phase fields $\phi_{\nu}$ change under the parity (${\cal P}$:
R$\leftrightarrow$L), and the spin-reversal transformations (${\cal T}$:
$\uparrow\leftrightarrow\downarrow$) as\cite{theta_fields}
\begin{mathletters}
\begin{eqnarray}
 {\cal P}:&&\phi_{\sigma}\rightarrow-\phi_{\sigma},\ \ \
            \phi_{\rho}\rightarrow-\phi_{\rho}\\
 {\cal T}:&&\phi_{\sigma}\rightarrow-\phi_{\sigma}
\end{eqnarray}
\end{mathletters}
Thus operators have discrete symmetries as ${\cal P}={\cal T}=1$ for the
singlet ($\sqrt{2}\cos\sqrt{2}\phi_{\sigma}$), and ${\cal P}={\cal
T}=-1$ for the triplet with $S^z=0$
($\sqrt{2}\sin\sqrt{2}\phi_{\sigma}$). The discrete symmetries of the
wave functions of these excited states are determined by combinations of
those of the ground state and the operators. Further discussion for the
discrete symmetries will be given in the next section.

\subsection{Renormalization Group}
Next, we consider the renormalization ($g_{1\perp}\neq 0$).
By the change of the cut off $\alpha\rightarrow {\rm e}^{{\rm d}l}\alpha$,
the coupling constant $g_{1\perp}$ and $K_\sigma$ are renormalized
as\cite{Kosterlitz}
\begin{mathletters}
\begin{eqnarray}
 \frac{{\rm d}y_0(l)}{{\rm d}l}&=&-y_1^{\ 2}(l),\\
 \frac{{\rm d}y_1(l)}{{\rm d}l}&=&-y_0(l) y_1(l),
\end{eqnarray}
\label{eqn:RGE}
\end{mathletters} 
where $y_0(l)\equiv 2(K_{\sigma}-1), y_1(l)\equiv g_{1\perp}/\pi v_{\sigma}$.
For the SU(2) symmetric case $y_{0}(l)=y_{1}(l)$
(the level-$1$ SU(2) Wess-Zumino-Novikov-Witten (WZNW)
model\cite{Knizhnik-Z,Gepner-W,Affleck-G-S-Z}) and $y_0(l)>0$, the
scaling dimensions of the operators for singlet and triplet excitations
split logarithmically by the marginally irrelevant coupling
as\cite{Giamarchi-S,Affleck-G-S-Z} (see Appendix \ref{apx:derivation})
\begin{mathletters}\label{eqn:sclngdim}
 \begin{eqnarray}
x_{\sigma}^{\rm singlet}&=&\frac{1}{2}+\frac{3}{4}\frac{y_0}{y_0\ln L+1},\\
x_{\sigma}^{\rm triplet}&=&\frac{1}{2}-\frac{1}{4}\frac{y_0}{y_0\ln L+1},
 \end{eqnarray}
\end{mathletters}
where $y_0\equiv y_0(0)$ and we have set $l=\ln L$. When $y_0<0$,
$y_0(l)$ is renormalized to $y_0(l)\rightarrow -\infty$, then a spin gap
appears. At the critical point ($y_0=0$), there are no logarithmic
corrections in the excitation gaps (the logarithmic correction from
higher order also vanish). Therefore, the critical point is obtained
from the intersection of the singlet and the triplet excitation
spectra\cite{Affleck-G-S-Z,Ziman-S,Okamoto-N}. In this case, we can
determine the critical point with high precision\cite{Okamoto-N}, since
the remaining correction is only $x_{\nu}=4$ irrelevant
fields\cite{Cardy86,Reinicke}. This irrelevant field, which does not
exist in the pure sine-Gordon model, comes from the nonlinear term
neglected when linearizing the dispersion relation near the Fermi level
in the course of the bosonization.

The physical meaning of this transition point ($y_0=0$) is the one where
the backward scattering coupling changes from repulsive to
attractive. Moreover, at the critical point, the SU(2) symmetry is
enhanced to the chiral SU(2)$\times$SU(2) symmetry\cite{Affleck-G-S-Z},
since the spin degrees of freedom of the right and the left Fermi points 
become independent.

Eq.(\ref{eqn:sclngdim}) also explains the fact that the SDW (CDW)
correlation is dominant for $K_{\rho}<1$ region with(out) spin gap,
while for $K_{\rho}>1$, the TS (SS) correlation is dominant with(out)
spin gap\cite{Giamarchi-S}.

Finally, let us consider the massive region. The behavior of the gap is
explained from the two-loop renormalization group equation of the
level-1 SU(2) WZNW model\cite{Amit-G-G,Destri,Nomura}
\begin{equation}
 \frac{{\rm d}y_0(l)}{{\rm d}l}=-y_0^{\ 2}(l)-\frac{1}{2}y_0^{\ 3}(l).
\end{equation}
If we define the correlation length $\xi$ as $y_0(\ln\xi)\equiv-1$ and
the energy gap as $\Delta E=v_{\sigma}/\xi$, then one can derive the
asymptotic form of the spin gap by solving the differential equation for
$|y_0(l)|\ll 1$ as
\begin{equation}
 \Delta E\propto\sqrt{|y_{0}|}\exp(-\mbox{Const.}/|y_0|).
  \label{eqn:gap}
\end{equation}
Note that eq.(\ref{eqn:gap}) is the same asymptotic behavior as the spin
gap of the negative-$U$ Hubbard model at half-filling given by
eq.(\ref{eqn:gap_Hubbard}).

\section{Uniqueness of Ground State and Discrete Symmetries}
\label{sec:symmetry}

In the previous section, the ground state is assumed to be a singlet, so 
that we should consider the way to make the singlet ground state in the
finite-size systems. Furthermore, we also discuss the discrete
symmetries of wave functions to identify the energy
spectra\cite{technical_issue}.  The symmetries depend on the choice of
representations for wave functions, so that we consider in the
representative two cases: one is the standard electron systems such as
the (extended) Hubbard model. The other is doped spin chains like the
$t$-$J$ model. In the following argument, the electron hopping is
restricted to the nearest neighbor, and the number of electrons is
assumed to be even. The results are summarized in
Table \ref{fig:symmetries}.

\subsection{Hubbard-type Models}

It is convenient to use the following representation of the basis to
describe the Hubbard-type models which permits double occupancy:
\begin{eqnarray}
 |\Psi_{\rm A}\rangle&\equiv&\sum_{n_1<\cdots<n_M; n_{M+1}<\cdots<n_N}
 \hspace{-1cm}f_{\rm A}(n_1,\cdots,n_M;n_{M+1},\cdots,n_N)
 \nonumber\\
 &&\times
 \prod_{i=1}^M   c^{\dag}_{n_i\downarrow}
 \prod_{j=M+1}^N c^{\dag}_{n_j\uparrow}|{\rm vac}\rangle,
\label{eqn:basis_A}
\end{eqnarray}

where $1\leq n_i\leq L$ and the periodicity $n_{i}+L\rightarrow n_i$ is
assumed. $N$ is number of electrons, and $M$ is number of electrons with
down spins. We call this representation as ``basis A''. In the case of
the (extended) Hubbard model, all off-diagonal matrix elements of the
Hamiltonian arise from the hopping term. Then the sign of these elements
are negative as far as no electron hops across the boundary. When an
electron moves across the boundary, the sign of the hopping amplitude
changes depending on $M$ reflecting anti-commutation relations of the
Fermi operators. If the periodic boundary conditions are assumed, the
hopping amplitude at the boundary become $+t$ for $M=$ even case, and
$-t$ for $M=$ odd case.

According to the Lieb-Schultz-Mattis theorem\cite{Lieb-S-M} (the
Perron-Frobenius theorem), if all off-diagonal elements of a real
symmetric irreducible matrix are non-positive, then the all vector
elements for the lowest eigen value have the same sign. Therefore if the
lowest eigen state is degenerate, these states can not be
orthogonal. Thus the ground state is proved to be a singlet. In order to
realize this situation in the (extended) Hubbard model with basis A, we
choose anti-periodic boundary conditions for $M=$ even case, and
periodic boundary conditions for $M=$ odd case. Then all off-diagonal
matrix elements become non-positive and the ground state is proved to be
a singlet. This selection of the boundary conditions are equivalent to
those derived from the Bethe-ansatz result of the Hubbard model in the
strong-coupling limit\cite{Ogata-S}.

The $t$-$J$ model (\ref{eqn:t-J}) can also be described by using the
basis A. This Hamiltonian has off-diagonal matrix elements originated
from the exchange interaction, in addition to the hopping term. The
exchange process gives $-J/2$, where the negative sign arises from the
anti-commutation relations of fermions. Thus the all off-diagonal matrix
elements are non-positive if the boundary conditions are chosen as
discussed above. This situation does not change if the three-site term
is added. Therefore the ground state of the $t$-$J$ model is also proved
to be a singlet.

Now we define an operation for the parity transformation (space
inversion) as
\begin{eqnarray}
 \lefteqn{{\cal P}f_{\rm A}(n_1,\cdots,n_M;n_{M+1},\cdots,n_N)=}\nonumber\\
 && f_{\rm A}(\bar{n}_M,\cdots,\bar{n}_1;\bar{n}_N,\cdots,\bar{n}_{M+1}),
\label{eqn:def_P_A}
\end{eqnarray}
where $\bar{n}_j\equiv L+1-n_j$. The spin-reversal transformation can be
defined only for $M=N/2$ cases as
\begin{eqnarray}
 \lefteqn{{\cal T}f_{\rm A}(n_1,\cdots,n_M;n_{M+1},\cdots,n_N)=}\nonumber\\
 && f_{\rm A}(n_{M+1},\cdots,n_{N};n_1,\cdots,n_{M}).
\label{eqn:def_T_A}
\end{eqnarray}
The eigen values of the operator ${\cal P}$ and ${\cal T}$ can take $\pm
1$. Since the all vector elements of the ground-state wave function have
the same sign, the ground state discussed above satisfies ${\cal
P}={\cal T}= 1$ and $k=0$.

The discrete symmetries of wave functions for the excited states are
determined by combinations of those of the ground state and the
operators given by the bosonization argument.  Thus ${\cal P}={\cal
T}=1$ for the singlet ($\sqrt{2}\cos\sqrt{2}\phi_{\sigma}$), and ${\cal
P}={\cal T}=-1$ for the triplet with $S^z=0$
($\sqrt{2}\sin\sqrt{2}\phi_{\sigma}$).  On the contrary, for the triplet
state with $S^z=\pm 1$ ($\exp(\pm{\rm i}\sqrt{2}\theta_{\sigma})$), the
symmetries are the same as those of the ground state (${\cal P}=1$,
$k=0$), because the all off-diagonal matrix elements are
non-positive. In this case, the SU(2) symmetry seems to be broken, this
discrepancy of the parity in the triplet states is due to the definition
of eq.(\ref{eqn:def_P_A}) which does not imply the change of the sign
stems from the permutation of fermions in the ${\cal P},{\cal T}$
transformations.  If we take account of the anti-commutation relation of
the Fermi operator in these transformations, then we get ${\cal
P}'=(-1)^{M}{\cal P}$ and ${\cal T}'=(-1)^{N/2}{\cal T}$. Thus the SU(2)
symmetry is recovered.

\subsection{Doped Spin Chains}

On the other hand, the models with a constraint to eliminate doubly
occupied sites, such as the $t$-$J$(-$J'$) model, are obtained by doping
holes into $S=1/2$ spin chains. These models are described by the
``basis B'' which is defined by
\begin{equation}
 |\Psi_{\rm B}\rangle\equiv\sum_{n_1<\cdots<n_N}
 f_{\rm B}(n_1, s_1;\cdots;n_N, s_N)
 \prod_{i=1}^N c^{\dag}_{n_i s_i}|{\rm vac}\rangle.
\label{eqn:basis_B}
\end{equation}
Hereafter, we argue based on the $t$-$J$ model. In order to make the
off-diagonal matrix elements stem from the exchange process
non-positive, we introduce a new basis $f'$ with a sign factor as
\begin{equation}
 f_{\rm B}(\cdots)\equiv(-1)^{\sum_{j=1}^N j(s_j+1/2)}f_{\rm B}'(\cdots),
\label{eqn:redef_basis_B}
\end{equation}
where $s_j=\pm 1/2$. Note that $j$ denotes the coordinate of the
squeezed spin system\cite{Ogata-S}. If an electron moves across the
boundary, this sign factor changes as
\begin{equation}
 (-1)^{\sum_{j=1}^N j(s_j+1/2)}
\rightarrow
 (-1)^{\sum_{j=1}^N j(s_j+1/2)}(-1)^M.
\label{eqn:negative_sign}
\end{equation}
Therefore, an additional negative sign appears if $M=$ odd. In addition
to this, a negative sign is also added, reflecting anti-commutation
relations of Fermi operators, so that the all off-diagonal matrix
elements become non-positive if we chose anti-periodic boundary
conditions for $M={\rm even}$, and periodic boundary conditions for
$M={\rm odd}$. Then all $f_{\rm B}'$ have the same sign so that the
ground state is proved to be a singlet. In this condition, the sign of
$f_{\rm B}$ does not change by the shift operation ($n_j\rightarrow
n_j+1$), so that the wave number of the ground state is $k=0$ for any
$M$.

Next, we define parity and spin-reversal transformations as follows:
\begin{eqnarray}
 {\cal P}f_{\rm B}(n_1, s_1;\cdots;n_N, s_N)&=&
 f_{\rm B}(\bar{n}_N, \bar{s}_1;\cdots;\bar{n}_1, \bar{s}_N),\\
 {\cal T}f_{\rm B}(n_1, s_1;\cdots;n_N, s_N)&=&
 f_{\rm B}(n_1, -s_1;\cdots;n_N, -s_N),
\end{eqnarray}
where $\bar{n}_j\equiv L+1-n_j$, $\bar{s}_j\equiv s_{N+1-j}$.  As we
have proved , all $f_{\rm B}'$ have the same sign in the ground state,
so that we only consider the variation of the sign factor of
eq.(\ref{eqn:redef_basis_B}) in these transformations. One can easily
show that the parity and the spin-reversal transformations bring
additional negative sign only for $M={\rm odd}$ case. Therefore, the
symmetries of the ground-state wave function are ${\cal P}={\cal T}=1$
for $N/2={\rm even}$, ${\cal P}={\cal T}=-1$ for $N/2={\rm odd}$.

The symmetries of the excited states can also be discussed in the same
way as in the case of the basis A.  In the basis B, ${\cal P}={\cal
T}=\pm 1$ for the singlet ($\sqrt{2}\cos\sqrt{2}\phi_{\sigma}$) and
${\cal P}={\cal T}=\mp 1$ for the triplet with $S^z=0$
($\sqrt{2}\sin\sqrt{2}\phi_{\sigma}$), where the upper (lower) sign
denotes $N/2=$ even (odd) case.  The triplet states with $S^z=\pm 1$ are
${\cal P}=\mp 1$ ($\exp(\pm{\rm i}\sqrt{2}\theta_{\sigma})$).  Note that
SU(2) symmetry is conserved in the parity of triplet excitations due to
the $M$ dependence of the parity.

For half-filling ($N=L$), the 1D $t$-$J$ model should be equivalent to
the $S=1/2$ Heisenberg spin chain.  The discrete symmetries known for
the Heisenberg chain are ${\cal P}={\cal T}=1$, $k=0$ for $L/2=$ even
and ${\cal P}={\cal T}=-1$, $k=\pi$ for $L/2=$ odd.  In this case, the
negative sign in eq.(\ref{eqn:negative_sign}) for $M=$ odd case can be
canceled by wave number $k=\pi$, instead of changing the boundary
conditions.  Therefore the symmetries in these two cases are consistent.

Unfortunately, the above proof can not be applied to the $t$-$J$-$J'$
model which includes the anti-ferromagnetic next-nearest-neighbor
interactions. However, boundary conditions and discrete symmetries of
this model are expected to be the same as those of the $t$-$J$ model, as
far as no instability takes place.

\section{Phase Diagrams of The Lattice Models}
\label{sec:diagrams}

Now we start our analysis for the models introduced in
Sec.\ref{sec:intro}. Besides the spin-gap instability, the charge
degrees of freedom is described by the single parameter $K_{\rho}$.  For
$K_{\rho}>1$, the superconducting correlations dominant, while for
$K_{\rho}\rightarrow\infty$, a phase separation takes place.  In order
to determine $K_{\rho}$, we need two independent physical quantities. In
our case, we calculate the compressibility and the Drude weight. In
finite-size systems, the compressibility is given by\cite{Ogata-L-S-A}
\begin{equation}
 \kappa=
\frac{L}{N^2}\left(\frac{E_0(L,N+2)+E_0(L,N-2)-2E_0(L,N)}{4}\right)^{-1}.
\label{eqn:kappa_num}
\end{equation}
where $E_0(L,N)$ is the ground state energy of a system with size $L$
and $N$ electrons ($n\equiv N/L$). On the other hand, the Drude weight
is given by the relation\cite{Kohn}
\begin{equation}
 D=\frac{L}{2}\left.\frac{\partial^2E_0(\Phi)}{\partial\Phi^2}\right|_{\Phi=0},
\end{equation}
where $\Phi$ is the flux which penetrates the ring. In the continuum
field theory, these two physical quantities are described by the
parameters of TL liquids\cite{Schulz,Kawakami-Y90a,Frahm-K}. The
compressibility is given as excitation $n_{\rho}=\pm 1$ in
eqs.(\ref{eqn:momentum}),(\ref{eqn:weight}):
\begin{equation}
 \frac{1}{n^2\kappa}=\frac{\pi}{2}\frac{v_{\rho}}{K_{\rho}},\label{eqn:kappa}
\end{equation}
and the Drude weight is given by the excitation $m_{\rho}=\Phi/\pi$ as
\begin{equation}
 D=\frac{K_{\rho}v_{\rho}}{\pi}.
\label{eqn:drude}
\end{equation}
Therefore, $K_{\rho}$ is obtained as $K_{\rho}=\pi\sqrt{Dn^2\kappa/2}$.

In order to obtain scaling dimensions of the spin degrees of freedom,
the spin-wave velocity is calculated by the following relation,
\begin{equation}
 v_{\sigma}=\lim_{L\rightarrow\infty}
  \frac{E(L,N,S=1,q=2\pi/L)-E_0(L,N)}{2\pi/L}.
\label{eqn:spin_velocity}
\end{equation}
The extrapolation is done by the function $v_{\sigma}(L) =
v_{\sigma}(\infty) + A/L^2 +B/L^4$.
These corrections are explained by the irrelevant fields with
$x_{\nu}=4$.

In the following, we analyze some models. Since there are too many
instabilities in the extended Hubbard model, we consider the $t$-$J$
model first to turn our attention on the spin-gap instability.

\subsection{The $t$-$J$ Model}
Here we analyze the spin-gap phase of the 1D $t$-$J$ model
(\ref{eqn:t-J}) by the above explained method. We diagonalize $L=8$-$30$
systems by the use of the Lanczos and the Householder algorism.

\begin{figure}[p]
 \begin{center}
 \epsfxsize=3.3in \leavevmode \epsfbox{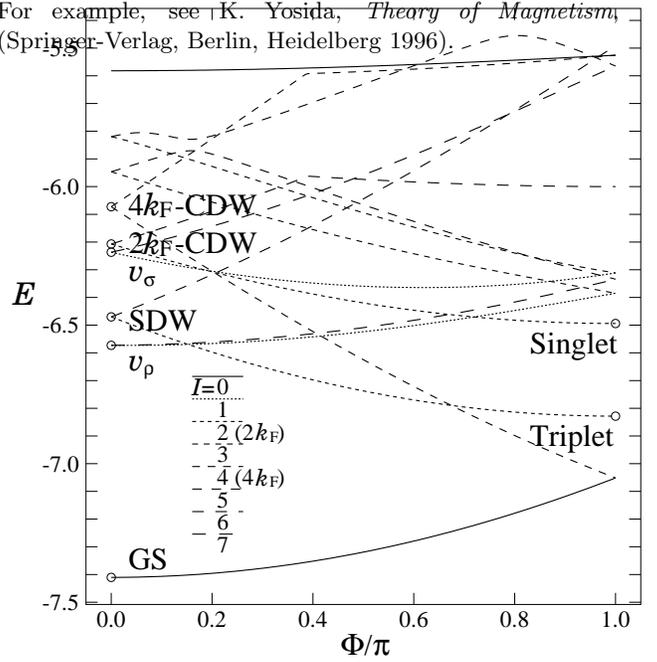}
 \end{center}
 \caption{Spectral flow of the 1D $t$-$J$ model at $J/t=2$ with length
 $L=8$ and electron number $N=4$. These energy spectra are the lowest
 two levels for each wave number $k=2\pi I/L$.
 The marked spectra correspond to the physical states written in this
 figure.}
 \label{fig:sflow}
\end{figure}

In order to investigate the structure of excitation spectra in detail,
we show in Fig.\ref{fig:sflow} the spectral flow (flux dependence of
energy) of $N/L=4/8$ system at $J/t=2$. In this case, the boundary
conditions are fixed to the ground state, so that the singlet and the
triplet excitation spectra appear at $\Phi=\pi$ which is equivalent to
the twisted boundary conditions. However, the wave number is not $k=0$
but $k=2k_{\rm F}$. This momentum shift is explained by the relation
$k(\Phi)=k(0)+N\Phi/L$\cite{momentum_shift}. The singlet and the triplet
excitation spectra are connected adiabatically from those of the CDW and
the SDW, respectively. If the linearized dispersion relation is exact,
these two spectra move parallelly versus $\Phi$ in this diagram. As
shown in eqs.(\ref{eqn:operators}), the charge degrees of freedoms
contribute the same amount in the CDW and the SDW spectra, so that the
critical point can be obtained by the level crossing of these
spectra. However, in this case, the precision become lower due to the
irrelevant fields, and the identification by the parity become
impossible due to the incommensurate wave number.

\begin{figure}
\epsfxsize=3.4in \leavevmode \epsfbox{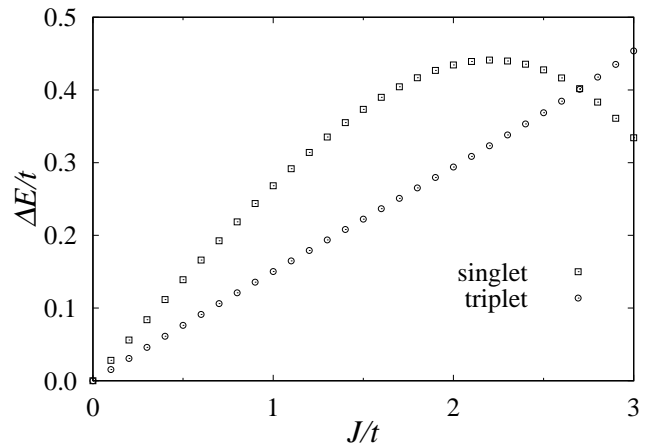}
\caption{Singlet and triplet excitation energies of the 1D $t$-$J$ model
 for $L=16$ system at $n=1/2$. These excitation spectra can be
 identified by symmetries of their wave functions.}
\label{fig:EXAMPLE}
\end{figure}

\begin{figure}
\epsfxsize=3.4in \leavevmode \epsfbox{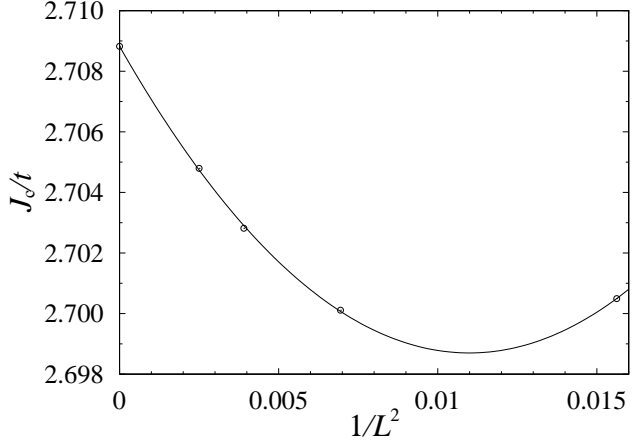}
\caption{Size dependence of $J_{\rm c}/t$ determined by the intersections
 of the excitation spectra for $L=8,12,16,20$ systems at $n=1/2$. These
 points are fitted by the form $A+B/L^2+C/L^4$.}
\label{fig:SIZE-QF}
\end{figure}

Figure \ref{fig:EXAMPLE} shows the singlet and the triplet excitation
spectra ($L=16,n=1/2$) versus $J/t$. The level crossing takes place at
$J/t\sim 2.7$. The size dependence of the critical point is shown in
Fig.\ref{fig:SIZE-QF}. Since the critical point is almost independent of
the system size, the phase diagram can be constructed without
extrapolation. Then we obtain Fig.\ref{fig:phsdgrm}(a).

In contrast to the former results\cite{Ogata-L-S-A,Hellberg-M,Chen-L},
the spin-gap phase spreads extensively toward the high-density
region. The spin-gap and phase-separation boundaries flow together
into the point $J/t\sim 3.5$ as $n\rightarrow 1$. We are not able to
answer whether the spin gap survives in the $n\rightarrow 1$ limit or
not, because the numerical results become unstable in the high density
region where the phase boundary is close to the phase-separated state.

In the low-density region, the phase boundary can be determined
analytically by solving a two-electron problem (see Appendix
\ref{apx:twoele}). Then the asymptotic behavior of the phase boundary
is obtained as
\begin{equation}
 2t/J_{\rm c}=\cos(\pi n/2).
\label{eqn:J_c-in-low-density}
\end{equation}
Note that the $J_{\rm c}$ given by eq.(\ref{eqn:J_c-in-low-density}) in
$L\rightarrow\infty$ limit is equivalent to the critical point where the
singlet pair forms a bound state in the ground state\cite{Lin}.  This
explains the fact that the spin-gap phase boundary overlaps the
$K_{\rho}=1$ line where the TL liquid behaves as free electrons, in the
low-density limit.

\begin{figure}
\epsfxsize=3.4in \leavevmode \epsfbox{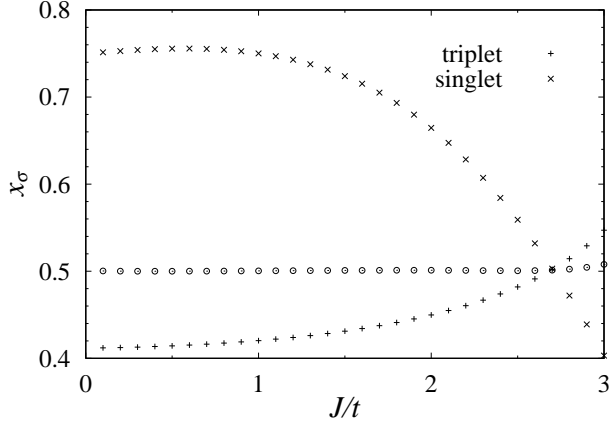}
\caption{Extrapolated value of
 $(x_{\sigma}^{\rm singlet}+3x_{\sigma}^{\rm triplet})/4$
 and the scaling dimensions for the singlet
 and the triplet excitations for $L=16$ system at $n=1/2$.}
\label{fig:RATIO}
\end{figure}

\begin{figure}
\epsfxsize=3.4in \leavevmode \epsfbox{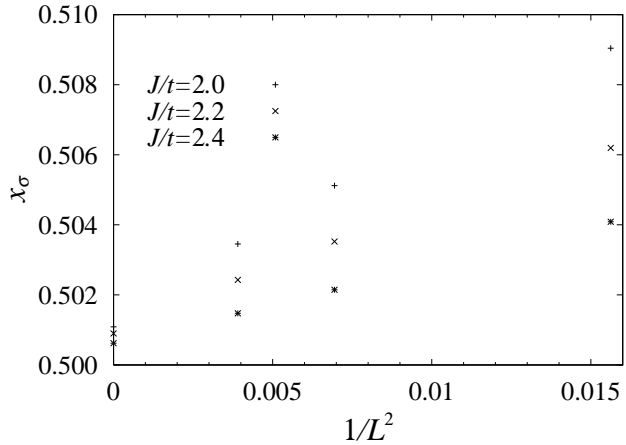}
\caption{Size dependence of the averaged scaling dimension
  $(x_{\sigma}^{\rm singlet}+3x_{\sigma}^{\rm triplet})/4$ at $n=1/2$.}
\label{fig:size_xs}
\end{figure}

In order to check the consistency of our argument, we calculate the
scaling dimensions for the singlet and the triplet excitations from
eqs.(\ref{eqn:energy}) and (\ref{eqn:spin_velocity}).
Then the average of the renormalized scaling dimension
(\ref{eqn:sclngdim}) is taken so as to eliminate the logarithmic
corrections as
\begin{equation}
\bar{x}_{\sigma}\equiv
\frac{x_{\sigma}^{\rm singlet}+3x_{\sigma}^{\rm triplet}}{4}.
\label{eqn:average}
\end{equation}
$\bar{x}_{\sigma}$ and its finite-size effect are shown in
Fig.\ref{fig:RATIO} and Fig.\ref{fig:size_xs}, respectively. The
extrapolated data become $1/2$ with error less than 0.2 \%.

In the spin-gap region ($J>J_{\rm c}$), the asymptotic behavior of the
spin gap is obtained using the relation $y_0\propto J-J_{\rm c}$ and
eq.(\ref{eqn:gap}).

\subsection{The $t$-$J$-$J'$ Model}
Next, we analyze the 1D $t$-$J$-$J'$ model (\ref{eqn:t-J-J'}).
According to Ref.\ref{Ogata-L-R}, the critical point for
the $J,J'\rightarrow 0$ limit is obtained by mapping the spin part onto
the case of $n=1$, using the factorized wave function,
\begin{mathletters}
\begin{eqnarray}
 J_{\rm eff}&=&J\langle n_i n_{i+1}\rangle_{\rm SF}+
 J'\langle n_i (1-n_{i+1})n_{i+2}\rangle_{\rm SF},\\
 J'_{\rm eff}&=&J'\langle n_i n_{i+1}n_{i+2}\rangle_{\rm SF},
\end{eqnarray}
\end{mathletters}
where $\langle\cdots\rangle_{\rm SF}$ indicates the expectation value of the
non-interacting spinless fermion. The effective ratio of the frustration
$\alpha_{\rm eff}$ is then obtained as
\begin{equation}
 \alpha_{\rm eff}(n,\alpha)=\left[
   \frac{(1+1/\alpha)n^{\,2}-s_2^{\,2}-s_1^{\,2}/\alpha}
   {n^3-(2s_1^{\,2}+s_2^{\,2})n+2s_1^{\,2}s_2}-1\right]^{-1},
  \label{eqn:eff_alpha}
\end{equation}
where $s_l\equiv \sin(l\pi n)/l\pi$. We can obtain the critical density
$n_{\rm c}$ where the spin gap vanishes, by comparing
eq.(\ref{eqn:eff_alpha}) with the result of the frustrated spin chain:
$\alpha_{\rm c}=0.2411$\cite{Okamoto-N}. For $\alpha=1/2$, we get
$n_{\rm c}=0.7433$.

On the other hand, in the low-density limit, the critical value for
the spin-gap phase $J_{\rm c}/t$, can be analytically obtained by
solving the two-electron problem as (see Appendix \ref{apx:twoele})
\begin{equation}
4t/J_{\rm c}=1+2\alpha+\sqrt{1+4\alpha^2}.
 \label{eqn:two_elec}
\end{equation}
The meaning of this point is same as that of the $t$-$J$ model.

\begin{figure}
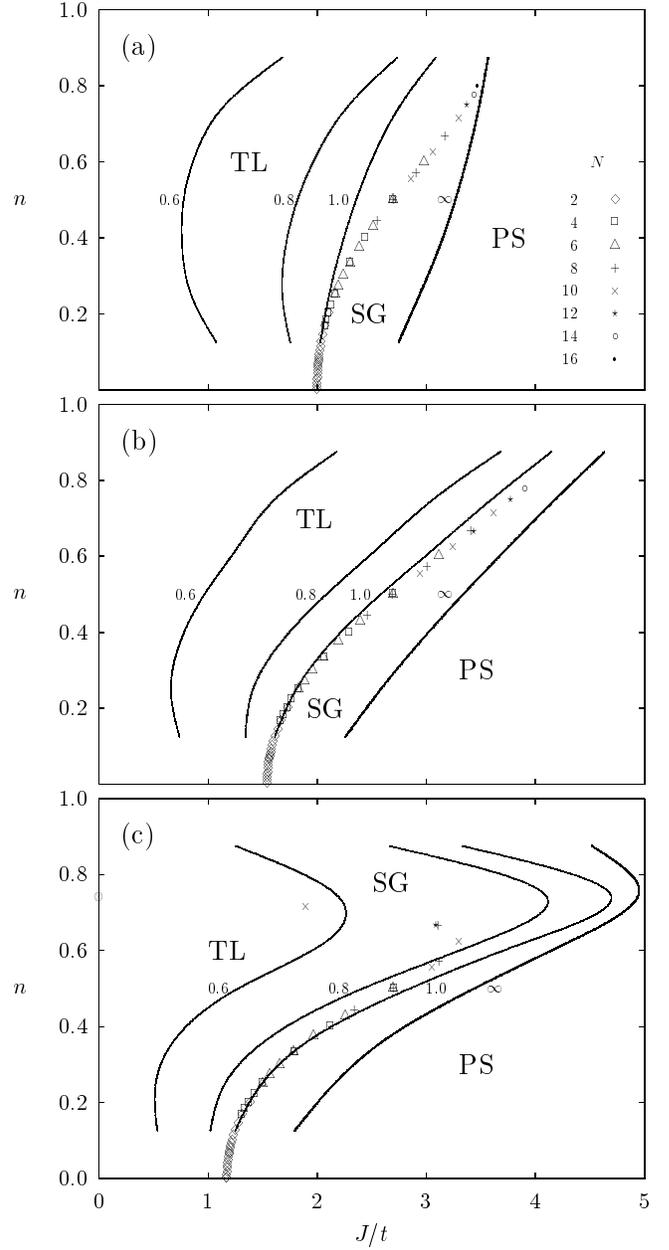

\noindent
 \epsfxsize=2.8in \leavevmode \epsfbox{pd_a000.epsi}\\
 \epsfxsize=2.8in \leavevmode \epsfbox{pd_a241.epsi}\\
 \epsfxsize=2.8in \leavevmode \epsfbox{pd_a500.epsi}\\
\caption{Phase diagrams of the 1D $t$-$J$-$J'$ model at
 (a) $\alpha=0$, (b) $\alpha=\alpha_{\rm c}$, (c) $\alpha=1/2$
  (TL: TL phase, SG: spin-gap phase, PS: phase-separated state).
 In the spin-gap phase where the backward scattering is attractive,
 the singlet excitation becomes lower than the triplet
 (see FIG.\protect{\ref{fig:EXAMPLE}},\protect{\ref{fig:RATIO}}).
 The contour lines of $K_{\rho}$ are calculated by the data of $L=16$
 system.}
 \label{fig:phsdgrm}
\end{figure}

We show the phase diagram of $\alpha=\alpha_{\rm c}$ case in
Fig.\ref{fig:phsdgrm}(b). The spin-gap phase boundary overlaps with the
contour line of $K_{\rho}=1$ at almost all densities. This situation is
quite resemble to that of the super-symmetric $t$-$J$ model with
long-range hopping and interactions\cite{Kuramoto-Y}. In this model,
there are no logarithmic corrections and the exact ground state is given
by the Gutzwiller wave function. This means that the charge degrees of
freedom is free electrons ($K_{\rho}=1$), and the singlet and the
triplet excitation spectra are degenerate for all densities.

Figure \ref{fig:phsdgrm}(c) is the phase diagram at $\alpha=1/2$. The
phase boundary starts from the critical value of the low-density limit
(\ref{eqn:two_elec}), and bends at $n\sim2/3$. It then flows into the
critical point $(J/t,n)=(0,n_{\rm c})$.  Thus the spin-gap phase with
different origin (the Majumder-Ghosh-like dimer phase in the low-doping
region, and the spin-gap phase in the large $J/t$ region) have a single
domain in the phase diagram. In contrast to the case of the $t$-$J$
model ($\alpha=0$), the spin-gap phase boundary lies in the $K_{\rho}<1$
region so that there is no TS region in this case.

Spin-gap phase may also exists for $\alpha>1/2$ cases. For example,
in $\alpha=\infty$ case, $J'_{\rm c}/t=1$ in the dilute limit, and
$n_{\rm c}=0.5752$ in the $J'\rightarrow 0$ limit.

In spite of the deformation of the phase diagram, the critical value
at the quarter-filling ($n=1/2$) is almost independent of the strength
of the frustration $\alpha$, and is kept at $J_{\rm c}/t\sim 2.7$.
Let us consider the reason for this using an argument based on the {\it
g-ology} model\cite{Emery,Solyom}. In order to apply the {\it g-ology}
model, we add the on-site Coulomb term ${\cal H}_U$ to
eq.(\ref{eqn:t-J-J'}) and relax the constraint. The original Hamiltonian
is restored when we set $U=\infty$.  Since the {\it g-ology} model is
appropriate for the weak coupling case, we consider $J'$ terms as
corrections to the $t$-$J$ model which belongs to the universality class
of the TL model. Then their contributions to the $g$-parameters, which
are related to the spin-gap generation, are identified as
\begin{equation}
 \delta g_{1\perp}=\delta g_{\sigma}
  =-J'(1+\cos 4 k_{\rm F}).
  \label{eqn:correction}
\end{equation}
For the quarter-filling, eq.(\ref{eqn:correction}) vanishes, so that the
$J'$ terms do not affect the renormalization flow of the spin part. Thus
the frustration does not change the critical point at the
quarter-filling within the scheme of the {\it g-ology} model.

\subsection{The Extended Hubbard Model}

The instability of the extended Hubbard model can be argued based on
the {\it g-ology} model for weak coupling cases ($U,V\rightarrow
0$)\cite{Emery,Solyom}. The $g$ parameters which are used to determine
the phase diagrams are identified as follows:
\begin{mathletters}
\begin{eqnarray}
 g_{1\perp}=g_{\sigma}&=&U+2V\cos 2k_{\rm F},\label{eqn:g_sigma}\\
 g_{\rho}&=&U+2V(2-\cos 2k_{\rm F}).\label{eqn:g_rho}
\end{eqnarray}
\end{mathletters}
We have defined $g_{\nu}\equiv g_{1\parallel}-g_{2\parallel}\mp
g_{2\perp}$ where the upper (lower) sign corresponds to $\nu=\rho$
($\nu=\sigma$). Then, from the discussion given in
Sec.\ref{sec:formalism}, the spin-gap phase boundary is determined by
$g_{1\perp}=g_{\sigma}=0$. On the other hand, the contour line for
$K_{\rho}=1$ is determined by the condition $g_{\rho}=0$, due to the
following relations
\begin{equation}
 K_{\nu}=\sqrt{\frac{2\pi v_{\rm F}+g_{4\parallel}\pm g_{4\perp}+g_{\nu}}
              {2\pi v_{\rm F}+g_{4\parallel}\pm g_{4\perp}-g_{\nu}}},
\end{equation}
where $g_{4\perp}=U+2V$, $g_{4\parallel}=2V$.

\begin{figure}
\noindent
\epsfxsize=3.4in \leavevmode \epsfbox{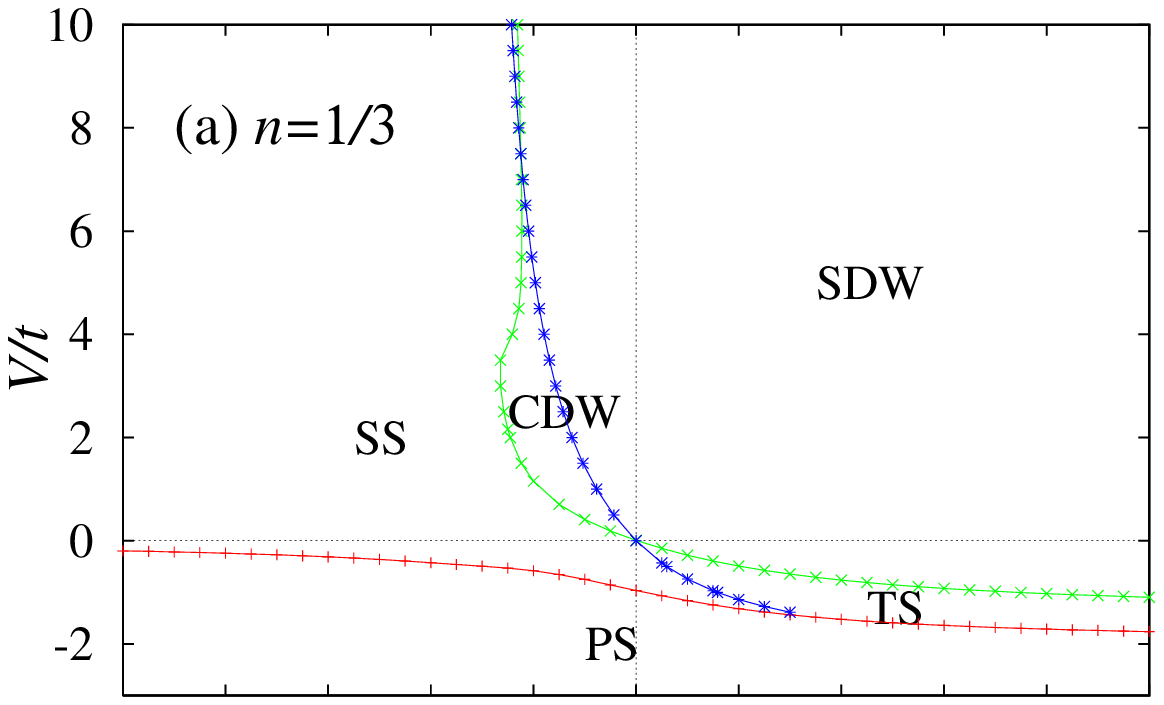}\\
\epsfxsize=3.4in \leavevmode \epsfbox{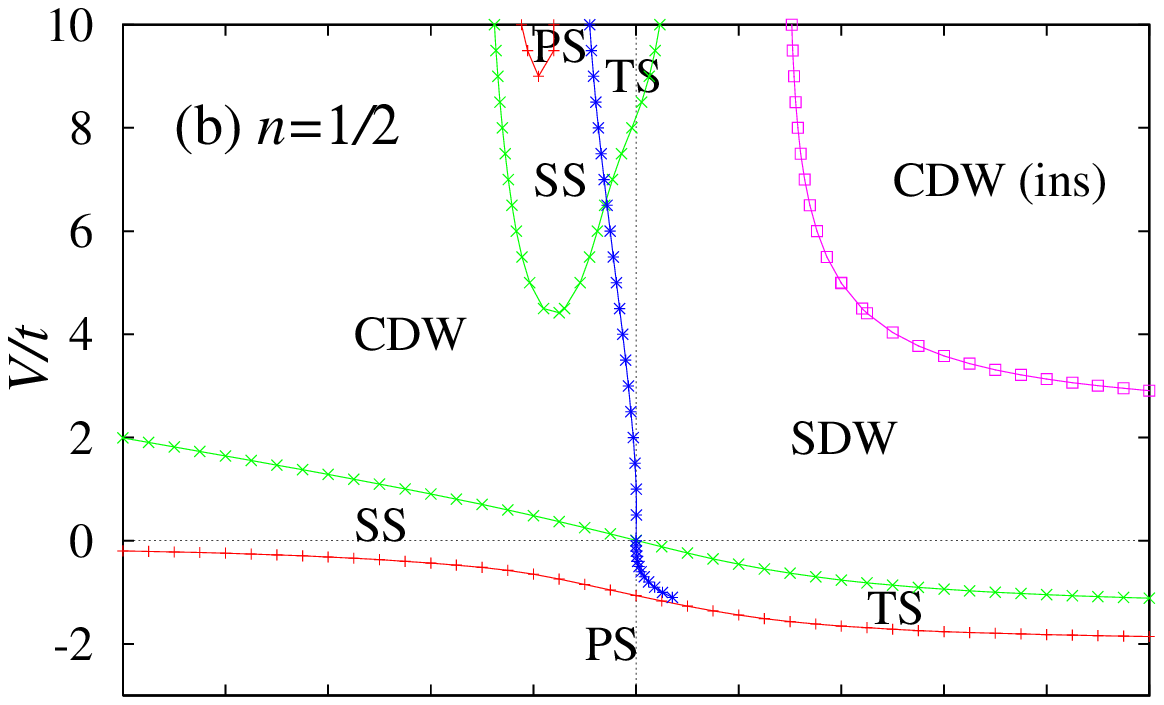}\\
\epsfxsize=3.4in \leavevmode \epsfbox{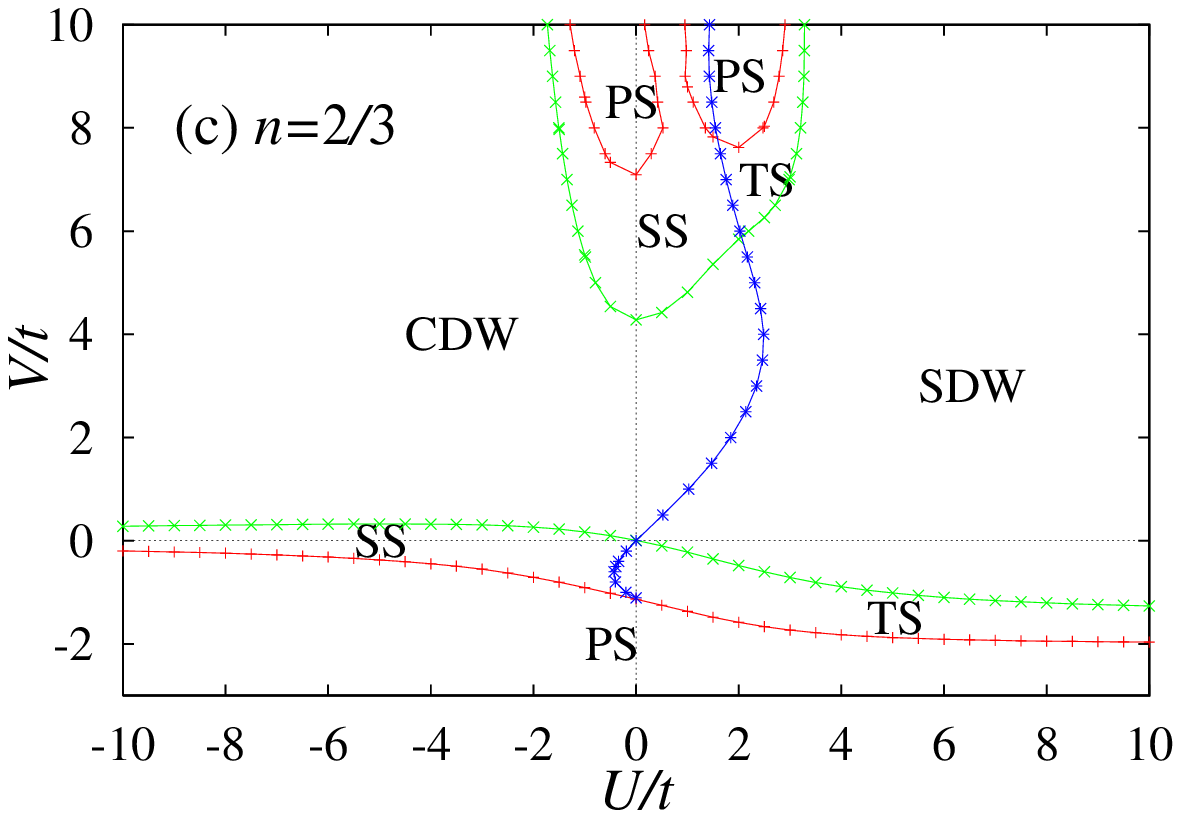}
\caption{Phase diagrams of the 1D extended Hubbard model determined by
 the data of $L=12$ systems at (a) $n=1/3$, (b) $n=1/2$, (c) $n=2/3$
  (SDW (TS): TL liquid phase with $K_{\rho}<1$ ($K_{\rho}>1$), CDW (SS):
 spin-gap phase with $K_{\rho}<1$ ($K_{\rho}>1$), PS: phase-separated
 state).}
\label{fig:tUV}
\end{figure}

Figure \ref{fig:tUV} shows the phase diagrams of the 1D extended Hubbard
model (\ref{eqn:EHM}) for various electron fillings. They are obtained
by analyzing the data of $L=12$ systems. In the all cases, the slopes of
the spin-gap phase boundaries and the $K_{\rho}=1$ contour lines near
the origin of the $U$-$V$ plain, are consistent with those predicted by
the {\it g-ology}. For $V<0$ region, there is phase-separated state and
its boundaries flow into $(U,V)=(\infty,-2t)$ due to the equivalence of
the XXZ spin chain in the large-$U$ limit\cite{Schulz}. The spin-gap
phase boundaries flow into these phase-separation boundaries.

In $n=1/3$ case, the spin-gap phase boundary and the $K_{\rho}=1$
contour line almost overlap near the solution of the two electron
problem (see Appendix \ref{apx:twoele}),
\begin{equation}
 V_{\rm c}=-\frac{2U_{\rm c}}{U_{\rm c}/t+4}.\label{eqn:two_Vc_tUV}
\end{equation}
This phenomenon is same as that of the low-density region of the
$t$-$J$(-$J'$) model.

At $n=1/2$, the spin-gap phase boundary is close to $U=0$, because the
effect of $V$ is canceled in eq.(\ref{eqn:g_sigma}).  In this phase
diagram, there are two regions with $K_{\rho}>1$.  Besides of the
spin-gap phase, a charge-gap phase exists for $U,V>0$ region due to the
Umklapp scattering. The analysis for this instability will be reported
elsewhere\cite{Nakamura_98}.

\begin{figure}
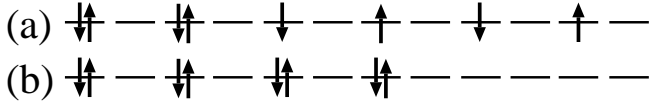

\noindent
\epsfxsize=3.4in \leavevmode \epsfbox{ps1.eps}\\ 
\epsfxsize=3.4in \leavevmode \epsfbox{ps2.eps}
\caption{Two phase-separated states which appear in the $V/t\gg 1$
 region of Fig.\protect{\ref{fig:tUV}}(c). (a) is located in $U/t>0$
 side. (b) lies in $U/t<0$ side. Spin-gap phase boundary exists between
 these two phase-separated states in the phase diagram.}
\label{fig:ps_fig}
\end{figure}

For $n=2/3$, a spin-gap phase appears in $U/t,V/t>0$
region\cite{Sano-O}. This is because the strong nearest-neighbor
repulsion stabilizes the on-cite singlet pairs. The one of the striking
feature in this phase diagram is that there are two phase separated
states in the $V/t\gg 1$ region, and the spin-gap phase boundary flows
between these two phase-separated states.  In this region, the spin-gap
phase boundary shifts to the large $U$ side due to the strong
finite-size effect.  The phase-separated state in the $U/t>0$ side is
considered as a mixture of $4k_{\rm F}$- and $2k_{\rm F}$-CDW
phases. The stability of this phase is already argued in
Ref.\ref{Sano-O} by using the second-order perturbation theory. On the
other hand, in the $U/t<0$ side, the system is separated into a $2k_{\rm
F}$-CDW phase and a vacuum. These phase-separated states are illustrated
in Fig.\ref{fig:ps_fig}.

\begin{figure}
\noindent
\epsfxsize=3.4in \leavevmode \epsfbox{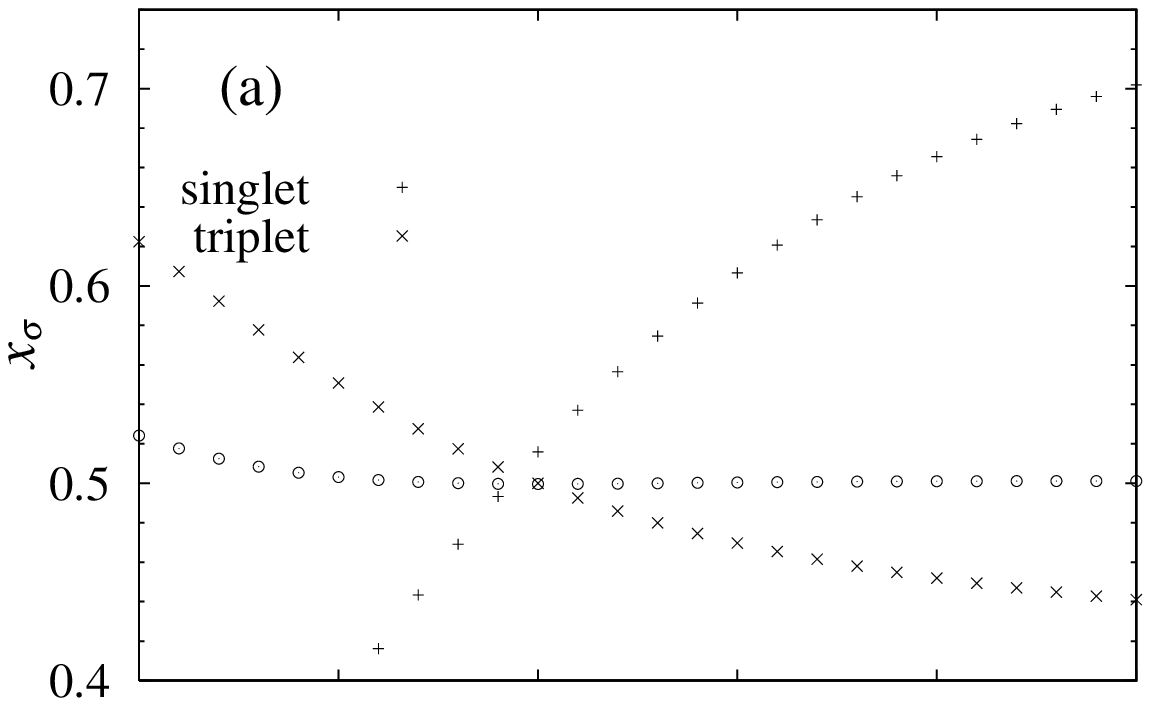}\\
\epsfxsize=3.4in \leavevmode \epsfbox{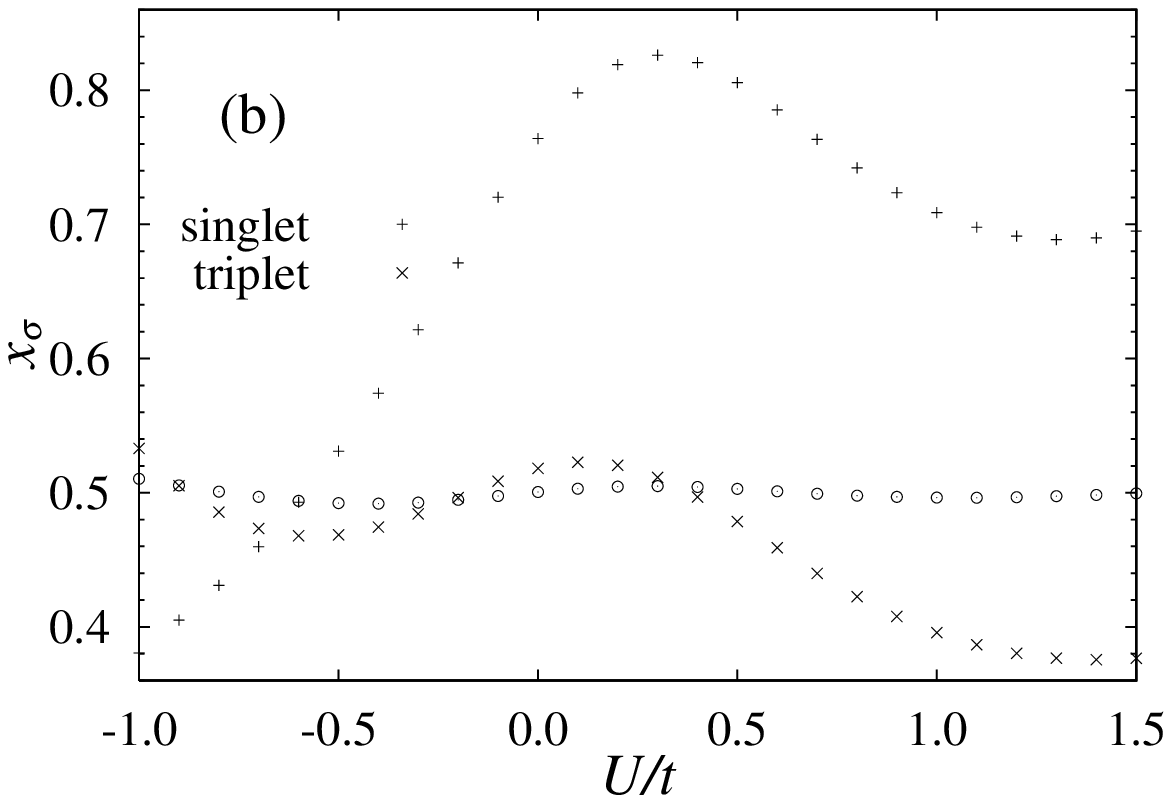}
\caption{Extrapolated value of $(x_{\sigma}^{\rm
singlet}+3x_{\sigma}^{\rm triplet})/4$ and the scaling dimensions for
the singlet and the triplet excitations for $L=16$ system at
$n=1/2$. (a) is the case of $V/t=2$, and (b) is the case of
$V/t=8$.}\label{fig:RATIO_2}
\end{figure}

The consistency of the argument can also be checked as in the case of
the $t$-$J$ model. Fig.\ref{fig:RATIO_2} shows the averaged scaling
dimension (\ref{eqn:average}) at $n=1/2$ for $V/t=2$ and $8$ cases,
calculated by the data of $L=8,12,16$ systems. Although the finite-size
effect is large for the $V\gg 1$ region, the extrapolated value become
$1/2$.

\section{Conclusion}\label{sec:summary}

To conclude, we have studied critical properties of spin-gap phases in
1D electron systems, considering the effect of the backward scattering
in TL liquids by the renormalization group analysis. The phase boundary
between TL liquids and spin-gap phases is shown to be determined by the
singlet-triplet level crossing point.  These excitation spectra are
extracted by twisting boundary conditions, and identified by the
discrete symmetries of wave functions. For this purpose, we have
discussed symmetries of wave functions under parity and spin-reversal
transformations.  We have applied the analysis to the extended Hubbard
model and the $t$-$J$-($J'$) model, and clarified the spin-gap regions
in the phase diagrams. The consistency of the our result has been
checked by investigating the ratio of logarithmic corrections.  Our
results are also consistent with those of the {\it g-ology} model in the
weak coupling limit, and of the two-electron problem in the dilute
limit.

\section{Acknowledgments}
M. N. thanks E. Dagotto, K. Itoh, K. Kusakabe, M. Ogata, K. Okamoto,
M. Oshikawa, H. Shiba, M. Takahashi, and H. Yokoyama for valuable
comments. A. K. is supported by JSPS Research Fellowships for Young
Scientists. The computation in this work was partly done using the
facilities of the Supercomputer Center, Institute for Solid State
Physics, University of Tokyo.

\appendix

\section{Quantum numbers in two notations}
\label{apx:woynarowich} In the analysis of 1D electron systems by
Bethe-ansatz results with CFT, a different notation from ours is often
used to describe the quantum
numbers\cite{Bares-B-O,Kawakami-Y90b,Woynarovich,Kawakami-Y90a,Frahm-K}.
In these notation the spin degrees of freedom is imposed only on down
spins. In their definition, $\Delta N_{\rm c}$ is the change of the
total number of electrons, and $\Delta N_{\rm s}$ is the change of the
number of down spins. $D_{\rm c}$ ($D_{ \rm s}$) denotes the number of
particles moved from the left charge (spin) Fermi point to the right
one. They are given by the eigen value of the number operator
$\hat{N}_{r,s}$ as
\begin{mathletters}
\begin{eqnarray}
  \Delta N_{\rm c} &=&
  N_{{\rm R}\uparrow}+N_{{\rm L}\uparrow}
 +N_{{\rm R}\downarrow}+N_{{\rm L}\downarrow},\\
 \Delta N_{\rm s} &=& N_{{\rm R}\downarrow}+N_{{\rm L}\downarrow},\\
  2D_{\rm c}      &=&  N_{{\rm R}\uparrow}-N_{{\rm L}\uparrow},\\
  2D_{\rm c} + 2D_{\rm s} &=&  N_{{\rm R}\downarrow}-N_{{\rm L}\downarrow}.
\end{eqnarray}\label{eqn:woynarovich}
\end{mathletters}
From eqs.(\ref{eqn:q_num}) and (\ref{eqn:woynarovich}), the quantum
numbers can be read as
$n_{\rho}=\Delta N_{\rm c}/2,
 n_{\sigma}=\Delta N_{\rm c}/2-\Delta N_{\rm s},
 m_{\rho}=2D_{\rm c}+D_{\rm s},
 m_{\sigma}=-D_{\rm s}$.
One can also easily show the equivalence of the selection rule
given by eq.(\ref{eqn:sel_rule}) and the one written by this
notation\cite{Woynarovich}:
\begin{mathletters}
\begin{eqnarray}
 D_{\rm c}&=&\frac{\Delta N_{\rm c}+\Delta N_{\rm s}}{2}
\ (\mbox{mod}\ 1),\\
 D_{\rm s}&=&\frac{\Delta N_{\rm c}}{2}\hspace{1.3cm} (\mbox{mod}\ 1).
\end{eqnarray}
\end{mathletters}
This relation is derived from the $U\rightarrow\infty$ limit of the
Hubbard model.

\section{Derivation of logarithmic corrections}
\label{apx:derivation}
Here we derive the logarithmic corrections given in
eq.(\ref{eqn:sclngdim}). Hereafter, we omit the spin index $\sigma$. We
consider
perturbation terms which break the scale invariance as
\begin{equation}
 {\cal H}
  ={\cal H}^{*}
  -\sum_{i}\int_0^L \frac{{\rm d}r}{2\pi}\lambda_i{\cal O}_i(r).
\end{equation}
Then the correction to the finite-size scaling is calculated within the
first-order perturbation as \cite{Cardy86}
\begin{eqnarray}
 E_i-E_0&=&
  \frac{2\pi v}{L}x_i-\sum_j\int_0^L\frac{{\rm d}r}{2\pi}\lambda_j
  \langle \phi_i|{\cal O}_j(r)|\phi_i\rangle\nonumber\\
  &=&
  \frac{2\pi v}{L}\left[x_i-\sum_j \lambda_j C_{iij}
		    \left(\frac{2\pi}{L}\right)^{x_j-2}\right],
		     \label{eqn:fss_correc}
\end{eqnarray}
where $|\phi_i\rangle$ is the eigen state of $E_i$, and $C_{ijk}$ is a
universal constant (OPE (operator product expansion) coefficient) fixed
by a three-point function:
\begin{equation}
 \langle {\cal O}_i(r_1){\cal O}_j(r_2){\cal O}_k(r_3)\rangle
  =\frac{C_{ijk}}
{r_{12}^{\ x_i+x_j-x_k}r_{23}^{\ x_j+x_k-x_i}r_{31}^{\ x_k+x_i-x_j}}.
\end{equation}
This coefficient can be derived from the following two ways.

\subsection{Abelian Bosonization}
The Lagrangian density of the spin part of eq.(\ref{eqn:effHam}) (the
sine-Gordon model) is written as
\begin{equation}
 {\cal L}={\cal L}_0+{\cal L}_I
\end{equation}
with
\begin{mathletters}
 \begin{eqnarray}
 {\cal L}_0&=&\frac{1}{2\pi}
\left[(v^{-1}\partial_{\tau}\phi)^2
+(\partial_x\phi)^2\right],\\
 {\cal L}_I&=&\frac{\lambda_0}{2\pi\alpha^2}{\cal O}_0
             +\frac{\lambda_1}{2\pi\alpha^2}{\cal O}_1,
 \end{eqnarray}
\end{mathletters}
where the perturbation term ${\cal L}_I$ consists of the following two
parts: one is a part of Gaussian model which denotes the deviation from
the free case ($K=1$). The other is the cosine term which stems
from the backward scattering. They denote the effect of interaction
between the left and the right Fermi points, and are written in the
Euclidean space as
\begin{mathletters}
\label{eqn:marginal_ops}
\begin{eqnarray}
 {\cal O}_0&\equiv&-\alpha^2K^{-1}\left[
 (v^{-1}\partial_{\tau}\phi)^2
+(\partial_x\phi)^2\right],\\
 {\cal O}_1&\equiv&\sqrt{2}\cos\sqrt{8}\phi.
\end{eqnarray}
\label{eqn:marginal_ops}
\end{mathletters}
Their coupling constants are given by
\begin{equation}
 2\lambda_0\equiv y_0(l),\ \ \ 
 \sqrt{2}\lambda_1\equiv y_1(l).
\end{equation}
For the SU(2) symmetric case $y_{0}(l)=y_{1}(l)$, and $y_0(l)>0$, the
marginally irrelevant coupling is calculated from eq.(\ref{eqn:RGE}) as
\begin{equation}
 y_0(l)=\frac{y_0}{y_0\ln L+1},\label{eqn:r_coupling}
\end{equation}
where the the bare coupling is defined as $y_0\equiv y_0(0)$, and we
have set $l=\ln L$.

Now we consider the operators for the singlet and the triplet states as
\begin{mathletters}
 \begin{eqnarray}
 {\cal O}_2&\equiv&\sqrt{2}\cos\sqrt{2}\phi,\\
 {\cal O}_3&\equiv&\sqrt{2}\sin\sqrt{2}\phi,\\
 {\cal O}_4&\equiv&\exp(+{\rm i}\sqrt{2}\theta),
 \end{eqnarray}
\end{mathletters}
then the coefficients of their OPE with the marginal operators
(\ref{eqn:marginal_ops}) are obtained as
\begin{eqnarray}
 C_{220}&=&C_{330}=-\frac{K}{2},\ \ \
 C_{440}=\frac{1}{2K},\nonumber\\
 C_{221}&=&-C_{331}=\frac{1}{\sqrt{2}},\ \ \
 C_{441}=0.
  \label{eqn:OPE_coeff}
\end{eqnarray}
Thus the scaling dimensions of the operators for singlet and triplet
excitations are obtained from eqs.(\ref{eqn:fss_correc}),
(\ref{eqn:r_coupling}), and (\ref{eqn:OPE_coeff}). These are consistent
with the results obtained by Gimarchi and Schulz \cite{Giamarchi-S}.

\subsection{Non-Abelian Bosonization}
In the standard bosonization theory, systems are described in U(1)
symmetric form, so that the explicit SU(2) symmetry in spin degrees of
freedom is lost. In order to describe systems with higher symmetry, it
is desirable to perform the calculation defining current fields that
conserve the SU(2) symmetry.

In SU(2) symmetric case, the system is described by chiral SU(2) currents
that are defined as
\begin{equation}
 \bm{J}_r\equiv
  :\psi_{r,\alpha}^{\dag}\frac{\bm{\sigma}_{\alpha\beta}}{2}\psi_{r,\beta}:,
\end{equation}
where $\bm{\sigma}=[\sigma^1,\sigma^2,\sigma^3]$ are the Pauli matrices
and $r=({\rm R},{\rm L})$.  The chiral SU(2) current $\bm{J}_{\rm R}$
has a conformal dimension $(\Delta^+,\Delta^-)=(1,0)$ and $\bm{J}_{\rm
L}$ has $(\Delta^+,\Delta^-)=(0,1)$.  The three components of
$\bm{J}_{r}$ obey commutation relations known as the Kac-Moody algebra
with central charge $k$:
\begin{equation}
J_r^i(z)J_r^j(w)=\frac{k/2}{(z-w)^2}\delta_{ij}
+\frac{{\rm i}\varepsilon_{ijl}\partial J_r^l(w)}{z-w}+\mbox{reg.},
\end{equation}
where $\varepsilon_{ijl}$ is the anti-symmetric structure factor.  For
spin-$s$ systems, there is a relation $k=s/2$. If a system is described
by this current algebra, the system belongs to the universality class of
the Wess-Zumino-Novikov-Witten non-linear $\sigma$ model with
topological coupling $k$ \cite{Knizhnik-Z,Gepner-W}.

In this case, the scaling dimension is
\begin{equation}
 x=\frac{2s_r(s_r+1)}{2+k},
\end{equation}
where $s_r=0,1/2,\cdots,k/2$.  Therefore, the lowest energy spectra for
the singlet and the triplet excitations are
\begin{equation}
 x^{\rm singlet}=x^{\rm triplet}=\frac{1}{2}.
\end{equation}

Now let us consider the correction in the presence of a marginal
operator ($x=2$)\cite{Affleck-G-S-Z} which is given by
\begin{equation}
 {\cal O}=\bm{J}_{\rm L}\cdot\bm{J}_{\rm R},
\end{equation}
The marginal operator ${\cal O}$ is proportional to $\bm{S}_{\rm
L}\cdot\bm{S}_{\rm R}$ where $\bm{S}_r$ is the SU(2) charge, and
$\bm{S}=\bm{S}_{\rm L}+\bm{S}_{\rm R}$ is the spin of the state
$\phi_i$. Letting the degrees of $\bm{S}$ and $\bm{S}_r$ are $s$ and
$s_r$, respectively, the expectation value becomes
\begin{equation}
 \langle \phi_i|\bm{S}_{\rm L}\cdot\bm{S}_{\rm R}|\phi_i\rangle
  =\frac{1}{2}(s(s+1)-s_{\rm L}(s_{\rm L}+1)-s_{\rm R}(s_{\rm R}+1)).
\end{equation}
Here, $s_{\rm L}=s_{\rm R}=1/2$ and $s=0$ for the singlet and $s=1$ for
the triplet. Thus the ratio of the logarithmic corrections is calculated
as $3:-1$.

\section{Dilute Limit}
\label{apx:twoele}
In the low-density limit, a many-body problem may be reduced to a
two-body problem. Here we consider a critical point where a bound
electron pair become stable in the ground state, and a singlet-triplet
level-crossing takes place. We perform the calculation following the
approach of H. Q. Lin, which was used for the 2D case\cite{Lin}.

In order to take the constraint of the $t$-$J$(-$J'$) model into
account, we relax the restriction, and add the on-site Coulomb term as
 \begin{equation}
  \tilde{{\cal H}}={\cal H}+U\sum_{i}n_{i\uparrow}n_{i\downarrow}.
 \end{equation}
The result of the original Hamiltonian can be obtained when we set
$U=\infty$ in the end of the calculation.

It is well known for a two-body problem that the ground state is a
singlet as far as the bottom of the energy band has no degeneracy
\cite{Slater-S-K,Yoshida}. This is consistent with the argument in
Sec.\ref{sec:symmetry}. The wave function in this system can be written
using the basis A as
\begin{equation}
 |\Psi\rangle=
 \sum_{ij}f(i,j)c_{i\uparrow}^{\dag}c_{j\downarrow}^{\dag}|{\rm vac}\rangle
\end{equation}
where $f(i,j)=f(j,i)$ for the singlet (${\cal T}=1$) as shown in
Sec.\ref{sec:symmetry}. The Schr\"{o}dinger equation for the singlet wave
function is
\begin{equation}
 Ef(i,j)=\sum_{l}[t_{il}f(l,j)+t_{jl}f(i,l)]
  +[U\delta_{ij}-J_{ij}]f(i,j),
 \label{eqn:S-eqn}
\end{equation}
with $t_{ij}=-t\delta_{|i-j|,1}$ and $J_{ij}=J\delta_{|i-j|,1}+\alpha
J\delta_{|i-j|,2}$ where $\alpha$ denotes the strength of the
frustration $\alpha\equiv J'/J$. The Fourier transformation of
eq.(\ref{eqn:S-eqn}) is given by
\begin{eqnarray}
Ef(k_1,k_2)&=&[t(k_1)+t(k_2)]f(k_1,k_2)\nonumber\\
&&+\frac{1}{L}\sum_{k}[U-J(k)]f(k_1+k,k_2-k),\label{eqn:Sch-eqn}
\end{eqnarray}
where
\begin{eqnarray}
 f(k_1,k_2)&=&
\frac{1}{L}\sum_{ij}f(i,j){\rm e}^{-{\rm i}k_1 r_i-{\rm i}k_2 r_j},\\
 t(k)&=&-2t\cos k,\\
 J(k)&=&2J(\cos k + \alpha \cos 2k).
\end{eqnarray}

Next, we introduce center of mass and relative momenta by
$Q=k_1+k_2, q=(k_1-k_2)/2$, and redefine the functions as
\begin{eqnarray}
 f_Q(k)&\equiv&f(k_1,k_2),\\
 \epsilon_Q(q)&\equiv&t(Q/2+q)+t(Q/2-q).
\end{eqnarray}
Then we get
\begin{equation}
 f_Q(q)=
  \frac{\frac{U}{L}\sum_{k}f_Q(k)-\frac{1}{L}\sum_{k}J(q-k)f_Q(k)}
  {E-\epsilon_Q(q)},
\end{equation}
where  
\begin{equation}
  J(q-k)=2J(\cos q \cos k + \alpha \cos 2 q \cos 2 k).\label{eqn:J_qk}
\end{equation}
Note that the terms that contain $\sin$ are omitted in eq.(\ref{eqn:J_qk}),
because they give no contribution due to their symmetry. Now we define
the following variables, and iterate them as
\begin{mathletters}\label{eqn:two}
 \begin{eqnarray}
 C_0&\equiv&\frac{U}{L}\sum_qf_Q(q)\nonumber\\
  &=&UI_{0,0}C_0-2JUI_{1,0} C_1-2\alpha JUI_{0,1} C_2,\\
 C_1&\equiv&\frac{U}{L}\sum_q f_Q(q)\cos q\nonumber\\
  &=&I_{1,0} C_0-2JI_{2,0} C_1-2\alpha JI_{1,1} C_2,\\
 C_2&\equiv&\frac{U}{L}\sum_q f_Q(q)\cos 2q\nonumber\\
  &=&I_{0,1} C_0-2JI_{1,1} C_1-2\alpha JI_{0,2} C_2,
 \end{eqnarray}
\end{mathletters}
where
\begin{equation}
 I_{m,n}\equiv\frac{1}{L}\sum_q \frac{\cos^m q\cos^n 2q}{E-\epsilon_Q(q)}.
\end{equation}
The criterion that eq.(\ref{eqn:two}) have a solution is
\begin{equation}
 \det\left[\begin{array}{ccc}
      1-UI_{0,0} & 2JUI_{1,0}     & 2\alpha JUI_{0,1}\\
      -I_{1,0}   & 2JI_{2,0}+1 & 2\alpha JI_{1,1}\\
      -I_{0,1}   & 2JI_{1,1}   & 2\alpha JI_{0,2}+1
 \end{array}\right]=0,
\label{eqn:equation_two}
\end{equation}
where all $I_{m,n}$ can be related to 
$I_{0,0}$ as follows,
\begin{eqnarray}
 I_{1,0}&=&(1-EI_{0,0})/4t,\nonumber\\
 I_{2,0}&=&-EI_{1,0}/4t,\nonumber\\
 I_{3,0}&=&(1-2EI_{2,0})/8t,\nonumber\\
 I_{4,0}&=&-EI_{3,0}/4t,\\
 I_{1,1}&=&2I_{3,0}-I_{1,0},\nonumber\\
 I_{0,1}&=&2I_{2,0}-I_{0,0},\nonumber\\
 I_{0,2}&=&I_{0,0}-4I_{2,0}+4I_{4,0},\nonumber
\end{eqnarray}
and $I_{0,0}$ diverges. Then setting $U=\infty$, we get the relation
between the singlet-state energy and the parameters of the model as
\begin{eqnarray}
4t/J&=&-z(4\alpha z^2-2\alpha+1)\nonumber\\
  &&+\sqrt{z^{2}(4\alpha z^2-2\alpha+1)^2-4\alpha(2z^{2}-1)},
\label{eqn:solution_tJJ_low}
\end{eqnarray}
where $z\equiv E/4t$. For the singlet pair with $Q=0$, the energy is given
by $E=-4t+B$ where $B$ is the binding energy. At the critical point
where the singlet pair becomes stable, the binding energy becomes
$B=0$. Then we get the solution (\ref{eqn:two_elec}) without size
dependence. In $\alpha=0$ case, we obtain $J_{\rm c}=2t$.

In the case of the extended Hubbard model, the solution can be obtained
by setting $(J,\alpha)=(-V,0)$ in eq.(\ref{eqn:equation_two}), and
leaving $U$ finite. The result is
\begin{equation}
 V=\frac{2U}{z(U/t-4z)}.
\label{eqn:two_sol_tUV}
\end{equation}
For $U\rightarrow 0$ limit, it coincides with the spin-gap phase
boundary and the $K_{\rho}=1$ contour line predicted by the {\it
g-ology}: $V=-U/2$.

Finally, we consider the singlet-triplet level-crossing point in the
dilute limit. In the system with anti-periodic boundary conditions, the
bottom of the energy band is degenerate, so that a level crossing may
take place. For the triplet state, the last term of
eq.(\ref{eqn:Sch-eqn}) vanishes due to the symmetry of the wave
function: $f(i,j)=-f(j,i)$ (${\cal T}=-1$). Therefore, the triplet state
is always non-interacting. This means that the level-crossing point can
be obtained as a solution (\ref{eqn:solution_tJJ_low}) for
$E=\epsilon_{Q=0}(\pi/L)$.  In this case, the density dependence of the
critical point of the $t$-$J$-$J'$ model can be expanded as
\begin{equation}
 J_{\rm c}(n)=J_{\rm c}(0)+A(\alpha)n^2+{\cal O}(n^4),
\end{equation}
where $J_{\rm c}(0)$ is same as the solution for the ground
state. Therefore, the spin-gap phase boundary in the low-density limit
coincides with the critical point for the bound electron pair in the
ground state, and its curve is the square-root type in the $J/t$-$n$
plane.  For $\alpha=0$ case, we obtain
eq.(\ref{eqn:J_c-in-low-density}). These solutions reflect the shape of
the band structure.

\begin{table}[h]
\begin{tabular}{c||rr|rr|rr}
      &\multicolumn{2}{c|}{basis A} &\multicolumn{2}{c|}{basis B} &&\\
\hline
      &${\cal P}$ &${\cal T}$ &${\cal P}$ &${\cal T}$ & $k$ &BC\\
\hline
 Ground state      &$ 1$ &$ 1$ &$\pm 1$ &$\pm 1$ &$ 0$ &$\mp 1$\\
 Singlet           &$ 1$ &$ 1$ &$\pm 1$ &$\pm 1$ &$ 0$ &$\pm 1$\\
 Triplet ($S^z=0$) &$-1$ &$-1$ &$\mp 1$ &$\mp 1$ &$ 0$ &$\pm 1$\\
 Triplet ($S^z=1$) &$ 1$ & $*$ &$\mp 1$ & $*$ &$ 0$ &$\pm 1$
\end{tabular}
\caption{Discrete symmetries of wave functions for different two bases
 (${\cal P}$: space inversion, ${\cal T}$: spin reversal, $k$:wave
 number BC: boundary conditions). The upper (lower) sign denotes the
 case of $N/2=$even (odd). These correspondences are explained by the
 Perron-Frobenius theorem and the bosonization theory.}
 \label{fig:symmetries}
\end{table}

\begin{references}
\bibitem{theta_fields}
 In these transformation, we fix the number of electrons and of spins
 ($n_{\nu}=$ Const.). Therefore, we do not consider the variation of
 $\theta_{\nu}$ fields.
\bibitem{technical_issue}
  For example, in the Lanczos algorism, energy spectra can be classified
 according to symmetries by choosing the initial vector.
\bibitem{momentum_shift}
 A system with twisted boundary conditions is transformed into a
 system with flux by the unitary operator
 $\exp(-{\rm i}\Phi\sum_{j=1}^L j n_j/L)$. In this transformation,
 the wave number $k$ in the former system shifts to $k+\Phi N/L$.
\end{references}
\end{document}